\documentclass[lettersize,journal]{IEEEtran}
\usepackage{amsmath,amsfonts}
\usepackage{algorithmic}
\usepackage{algorithm}
\usepackage{array}
\usepackage[caption=false,font=normalsize,labelfont=sf,textfont=sf]{subfig}
\usepackage{textcomp}
\usepackage{stfloats}
\usepackage{url}
\usepackage{verbatim}
\usepackage{graphicx}
\usepackage{cite}
\usepackage{diagbox}
\usepackage{cleveref}
\usepackage{multirow}
\usepackage[table]{xcolor}
\usepackage{cite}
\begin{document}

\title{An Enhanced Submodule for Modular Multilevel Converters With DC Fault-Clearing Capability}

\author{Yu Liu, Shengbo Zhang, and Yating Yuan

\thanks{Yu Liu and Shengbo Zhang are with the College 
	of Urban Transportation and Logistics, Shenzhen Technology University,
	Shenzhen, China.}
\thanks{Yating Yuan is with the Department of Applied Mathematics, University
	of Waterloo, 200 University Avenue, Waterloo, Ontario, Canada.}
\thanks{Manuscript received November 6, 2026. Corresponding authors: Yu Liu (liuyu@sztu.edu.cn) and Yating Yuan (yating.yuan@uwaterloo.ca).}}

\markboth{Journal of \LaTeX\ Class Files,~Vol.~14, No.~8, Novermber~2026}%
{Shell \MakeLowercase{\textit{et al.}}: A Sample Article Using IEEEtran.cls for IEEE Journals}


\maketitle

\begin{abstract}
Modular multilevel converter (MMC) has been successfully applied in various power electronic systems owing to its high efficiency, scalability, and superior output performance. Although the half-bridge submodule (HBSM) is widely used in MMCs for its structural simplicity, it is incapable of handling direct-current (DC) short-circuit faults. The diode-clamp submodule (DCSM) addresses this limitation by providing DC fault ride-through capability. However, because its two identical capacitors are connected in series, the equivalent capacitance is halved. To overcome this drawback, an enhanced SM is proposed in this paper. For the same energy storage capacity, the proposed SM reduces the total required capacitance by 75\% compared with the DCSM. In addition, the proposed SM requires one fewer diode than the DCSM, thereby lowering the overall MMC capital cost. The topology and the operating modes of the proposed SM are described in detail, and its functionality is experimentally validated. The results demonstrate that the proposed SM can suppress DC fault currents and restore normal operation without additional external protection devices.
\end{abstract}


\section{Introduction}
\IEEEPARstart{M}{odular} multilevel converter (MMC) has been successfully employed in various high-voltage direct current applications such as point-to-point systems \cite{yang2012} or multi-terminal systems \cite{li2014,Tang2018}. Moreover, it can be employed in flexible alternating current transmission system \cite{Sotoodeh2014}, static compensator \cite{Yu2016}, motor drives \cite{picas2018}, and medium- to low-voltage direct-current (DC) grids \cite{zhao2021}. Owing to its high modularity and scalability, high energy efficiency, high voltage rating, low harmonics distortion, and the absence of DC link \cite{debnath2015}, the technology has attracted considerable research interest. Consequently, numerous topologies and control methods have been proposed in the literature. 

In an MMC, the multilevel voltage waveform is generated by switching multiple submodules (SMs) connected in series. The half-bridge submodule (HBSM) is the first submodule topology proposed for MMC \cite{lesnicar2003}, and it has been most commonly used \cite{paez_overview_2019}. The HBSM contains only one capacitor and two switches, making it cost-effective and easy to control. However, it can only block current through the lower arm in one direction. Current in the opposite direction flows freely through the free-wheeling diode. Thus, the HBSM lacks the capability to interrupt short-circuit fault currents. To clear fault currents, additional devices such as circuit breakers must be installed on either the alternating current- (AC) or DC-side of the MMC. Notably, a circuit breaker typically requires several cycles to clear a fault, resulting in a slow fault clearance \cite{debnath2015}.

To endow the MMC with fault-handling capability, the diode-clamp submodule (DCSM) was proposed in \cite{li2013a}. Compared with an HBSM, the DCSM contains one extra switch, one extra capacitor, and two additional diodes. The extra switch is connected in anti-series with the lower-arm switch of the HBSM, enabling it to block current in the reverse direction and thus allowing the lower arm to block current in either direction. The extra capacitor is used to absorb power and provide voltage support when fault current flows in the reverse direction. However, the DCSM contains two identical capacitors connected in series, which reduces the equivalent capacitance to half of the individual capacitance.

Given that an MMC typically consists of hundreds or even thousands of SMs, any additional cost for a single component will result in a significantly higher capital cost for the entire MMC station. To reduce the capital cost of the MMC, an enhanced HBSM is proposed in this paper. Similar to the DCSM, the proposed SM can autonomously suppress DC fault currents and resume normal operation after fault clearance. More importantly, for the same energy storage capacity, the proposed SM requires a smaller total capacitance and one fewer diode than the DCSM, thereby reducing the overall capital cost of the MMC.

This paper is organized as follows: Section \ref{topology} presents the topology and operating modes of the proposed SM. Then, the functionality and the performance of the proposed SM is demonstrated through a physical experiment in section \ref{experiment}. Finally, conclusions are drawn in Section \ref{conclusion}.

\section{Topology and operating modes of the Proposed Submodule} \label{topology}
\begin{figure}[htb]
	\centering
	\includegraphics[width=4.5cm]{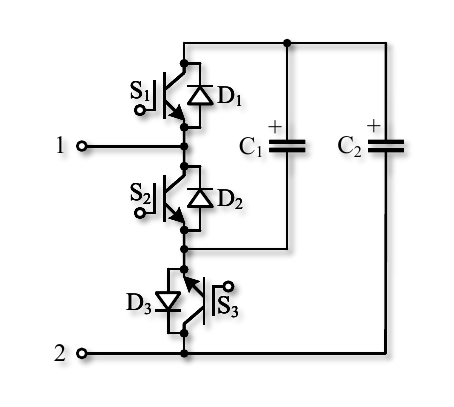}
	\caption{Circuit topology of the proposed SM.}
	\label{SM_Topology}
\end{figure}

To reduce the component count while maintaining fault-handling capability, a novel topology of an SM is proposed as the circuit diagram shown in Fig. \ref{SM_Topology}. The proposed SM is a two-port circuit network, which consists of three fully controlled switches $\rm S_1$ -- $\rm S_3$, three diodes $\rm D_1$ -- $\rm D_3$, and two capacitors $\rm C_1, C_2$ where $\rm C_1 > C_2$. The switches and diodes form a half-bridge configuration: the upper arm comprises switch $\rm S_1$ with an anti-parallel connected diode $\rm D_1$, while the lower arm consists of switches $\rm S_2$ and $\rm S_3$ connected in anti-series, each with an anti-parallel diode. The corresponding diodes in the lower arm are thus also in anti-series. The diodes serve to clamp the voltages across the switches and provide a free-wheeling path for current when the switches are turned off. Capacitor $\rm C_1$ is connected in parallel to the branch formed by $\rm S_1$ and $\rm S_2$, while capacitor $\rm C_2$ is connected in parallel with the entire half-bridge. Compared with a DCSM \cite{li2013a}, a proposed SM requires one fewer diode, thereby reducing the component count and overall system complexity for the MMC. 

The proposed SM features three operating modes similar to those of the conventional HBSM, namely, the inserted mode, the bypassed mode, and the blocked mode. During normal operation, the SM switches between the inserted and bypassed modes. Moreover, existing control methods suitable for the HBSM can also be directly applied to the proposed SM.

\subsection{Mode 1: Inserted Mode}
\begin{figure}[htb]
	\centering
	\vspace{-0.5cm}
	\includegraphics[width=9cm]{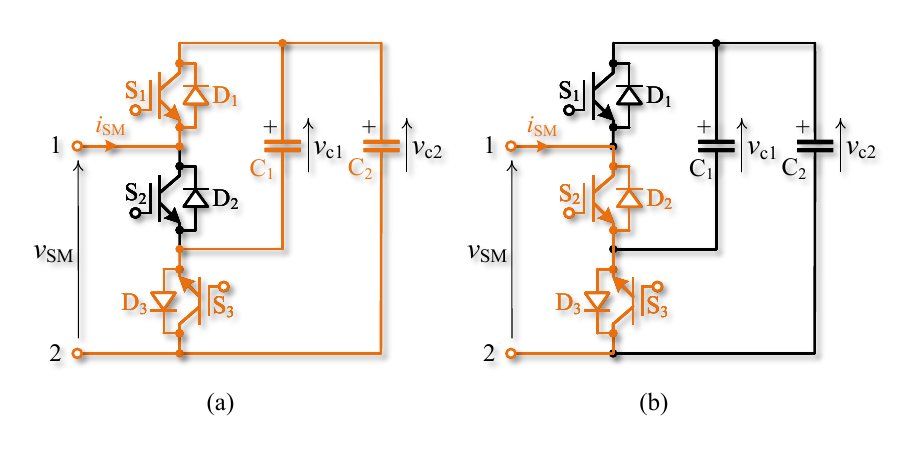}
	\caption{Current path in the SM during normal operation: (a) inserted mode; (b) bypassed mode.}
	\label{Operation_Normal}
\end{figure}

In this mode, $\mathrm{S}_1$ and $\mathrm{S}_3$ are turned on, whereas $\mathrm{S}_2$ is turned off. Consequently, the upper arm conducts while the lower arm is blocked, and both capacitors are connected in parallel and inserted into the external circuit. The current path in the SM is marked in orange, as shown in Fig.~\ref{Operation_Normal}(a). When current enters the SM at port 1 ($i_{\mathrm{SM}} > 0$), it flows through diode $\mathrm{D}_1$ and then splits into two branches: one branch goes through $\mathrm{C}_1$ and $\mathrm{D}_3$, while the other flows through $\mathrm{C}_2$. Finally, the two branches merge at port 2. During this process, the capacitors are charging, and their voltages $v_{\mathrm{c1}}$ and $v_{\mathrm{c2}}$ increase. 

Conversely, when current enters the SM from port 2 ($i_{\mathrm{SM}} < 0$), it also splits into two branches: one flows through $\mathrm{S}_3$ and $\mathrm{C}_1$, while the other flows through $\mathrm{C}_2$. Both paths then merge and flow through $\mathrm{S}_1$ to port 1. In this case, the capacitors are discharging, and $v_{\mathrm{c1}}$ and $v_{\mathrm{c2}}$ decrease. The voltage across the two ports, $v_{\mathrm{SM}}$, can be calculated as follows:
\begin{equation}
	\label{v_sm1}
	v_{\rm SM} = 
	\begin{cases}
		v_{\rm c1} + v_{\rm d1} + v_{\rm d3} = v_{\rm c2} + v_{\rm d1}, &\text{if $i_{\rm SM} > 0$}\\
		v_{\rm c1} + v_{\rm s1} + v_{\rm s3} = v_{\rm c2} + v_{\rm s1}, &\text{if $i_{\rm SM} < 0$}
	\end{cases},
\end{equation}
where $v_{\mathrm{s1}}$, $v_{\mathrm{s3}}$, $v_{\mathrm{d1}}$, and $v_{\mathrm{d3}}$ are the voltage drops across $\mathrm{S}_1$, $\mathrm{S}_3$, $\mathrm{D}_1$, and $\mathrm{D}_3$, respectively. During normal operation, these voltage drops are much smaller than the capacitor voltages, yielding $v_{\mathrm{SM}} \approx v_{\mathrm{c1}} \approx v_{\mathrm{c2}}$ in this mode. Therefore, the two capacitors are effectively connected in parallel regardless of the direction of $i_{\mathrm{SM}}$.

\subsection{Mode 2: Bypassed Mode}
When $\mathrm{S}_1$ is turned off and $\mathrm{S}_2$ and $\mathrm{S}_3$ are turned on, the upper arm is blocked while the lower arm conducts, and the SM operates in the bypassed mode. The current path is shown in orange in Fig.~\ref{Operation_Normal}(b). When $i_{\mathrm{SM}} > 0$, it flows directly through $\mathrm{S}_2$ and $\mathrm{D}_3$, and then exits the SM through port 2. When $i_{\mathrm{SM}} < 0$, it flows through $\mathrm{S}_3$ and $\mathrm{D}_2$, and then exits the SM through port 1. In both cases, $i_{\mathrm{SM}}$ does not flow through the capacitors. Hence, the capacitors are bypassed and their voltages remain constant, neglecting losses. In this mode, $\mathrm{C}_1$ and $\mathrm{C}_2$ are always connected in parallel regardless of the direction of $i_{\mathrm{SM}}$, and $v_{\mathrm{SM}}$ can be calculated as follows:
\begin{equation}
	v_{\rm SM} = v_{\rm s2} + v_{\rm d3} = v_{\rm s3} + v_{\rm d2} \approx 0.
\end{equation}

In Modes 1 and 2, $\mathrm{C}_1$ and $\mathrm{C}_2$ are connected in parallel. Thus, the equivalent capacitance of the proposed SM, denoted as $\mathrm{C}_{\mathrm{eq}}^{\mathrm{p}}$, is the sum of the capacitance of both capacitors, i.e.:
\begin{equation}
	\label{C_p}
	\mathrm{C}_{\mathrm{eq}}^{\mathrm{p}} = \mathrm{C}_1 + \mathrm{C}_2.
\end{equation}
This suggests that $\mathrm{C}_{\mathrm{eq}}^{\mathrm{p}}$ is greater than $\mathrm{C}_1$. By contrast, the two capacitors in the DCSM are connected in series and with the same capacitance \cite{li2013a}, i.e., $\mathrm{C}_1 = \mathrm{C}_2$, the equivalent capacitance of the DCSM, denoted as $\mathrm{C}_{\mathrm{eq}}^{\mathrm{D}}$, can be calculated as:
\begin{equation}
	\label{C_d}
	\mathrm{C}_{\mathrm{eq}}^{\mathrm{D}} = \frac{\mathrm{C_1}\mathrm{C_2}}{\mathrm{C_1} + \mathrm{C_2}} = \frac{1}{2}\mathrm{C}_1
\end{equation}
This shows that $\mathrm{C}_{\mathrm{eq}}^{\mathrm{D}}$ is just a half of $\mathrm{C}_1$. Therefore, for the same $\rm C_1$ and $\rm C_2$, $\mathrm{C}_{\mathrm{eq}}^{\mathrm{p}} > \mathrm{C}_{\mathrm{eq}}^{\mathrm{D}}$, indicating that one proposed SM offers greater energy storage capacity than one DCSM does. From another perspective, to offer both SMs the same energy storage capacity, i.e.,  $\mathrm{C}_{\mathrm{eq}}^{\mathrm{p}} = \mathrm{C}_{\mathrm{eq}}^{\mathrm{D}}$,
the total capacitance needed for a proposed SM is $\mathrm{C}_1 + \mathrm{C}_2 $, which is equal to $\mathrm{C}_{\mathrm{eq}}^{\mathrm{p}}$ according to \cref{C_p}. For the DCSM, however, $\mathrm{C}_1 = \mathrm{C}_2 = 2\mathrm{C}_{\mathrm{eq}}^{\mathrm{p}}$ according to \cref{C_d}, the total capacitance needed for a DCSM is thus $\mathrm{C}_1 + \mathrm{C}_2 = 4\mathrm{C}_{\mathrm{eq}}^{\mathrm{p}}$. Therefore, a proposed SM requires 75\% less capacitance compared to a DCSM for the same energy storage capacity, thereby reducing the building cost for MMC.

\subsection{Mode 3: Blocked Mode}
\begin{figure}
	\centering
	\vspace{-0.25cm}
	\includegraphics[width=9cm]{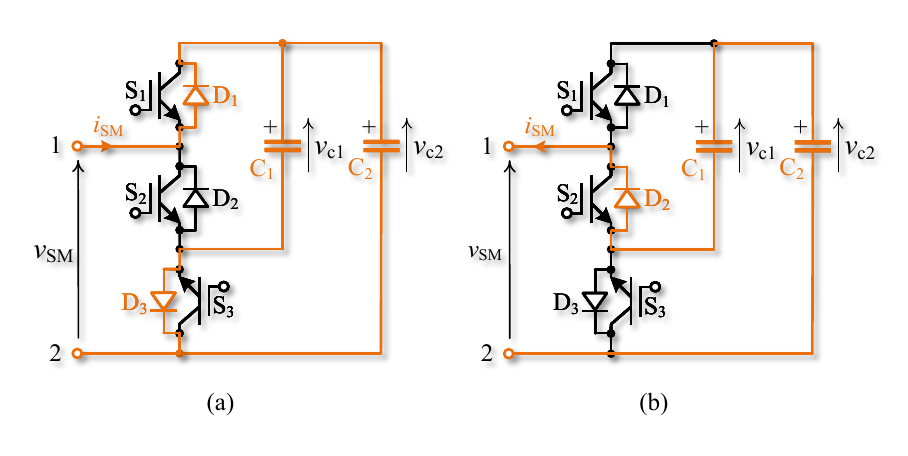}
	\vspace{-0.75cm}
	\caption{Current path in the blocked mode: (a) $i_{\rm SM} > 0$; (b) $i_{\rm SM} < 0$.}
	\label{Operation_Blocked}
\end{figure}

When all switches $\mathrm{S}_1$ -- $\mathrm{S}_3$ are turned off, the SM operates in the blocked mode. This mode can be adopted during the startup process to charge the capacitor voltages to the preset value, or when a short-circuit fault occurs. When $i_{\mathrm{SM}} > 0$, the current flows through $\mathrm{D}_1$ and then splits into two branches: one flows through $\mathrm{C}_1$ and $\mathrm{D}_3$, while the other flows through $\mathrm{C}_2$. Finally, the two branches merge at port 2. The current path is shown in orange in Fig.~\ref{Operation_Blocked}(a). In this case, the two capacitors are connected in parallel, and $v_{\mathrm{SM}}$ can be calculated as \cref{v_sm1}. If $i_{\mathrm{SM}}$ is a fault current, it will be suppressed because the boosting direction of $v_{\mathrm{SM}}$ opposes the direction of $i_{\mathrm{SM}}$ in this mode.

If $i_{\mathrm{SM}} < 0$, however, the current flows successively through $\mathrm{C}_2$, $\mathrm{C}_1$, and then $\mathrm{D}_2$ to port 1. The current path through the SM is shown in Fig.~\ref{Operation_Blocked}(b). Unlike the previous case, $\mathrm{C}_1$ and $\mathrm{C}_2$ are connected in series in this mode, with $v_{\mathrm{c1}}$ increasing while $v_{\mathrm{c2}}$ decreases. The voltage $v_{\mathrm{SM}}$ can then be calculated as follows:
\begin{equation}
	\label{v_SM3}
	v_{\rm SM} = v_{\rm c2} - v_{\rm c1} - v_{\rm d2}.
\end{equation}

\subsection{Fault Clearing}
In Modes 1 and 2, $v_{\mathrm{c1}} \approx v_{\mathrm{c2}}$. Thus, $v_{\mathrm{SM}}$ is initially zero when the SM transitions from either of these modes to Mode 3. Once blocked, $v_{\mathrm{c2}}$ decreases while $v_{\mathrm{c1}}$ increases because the two capacitors are connected in anti-series, resulting in a negative $v_{\mathrm{SM}}$ that continuously decreases. Since $\mathrm{C}_1 > \mathrm{C}_2$, $v_{\mathrm{c2}}$ decreases at a higher rate than $v_{\mathrm{c1}}$ increases. Consequently, $v_{\mathrm{SM}}$ becomes increasingly negative, establishing a voltage that opposes $i_{\mathrm{SM}}$ and thereby suppresses the fault current.

\begin{figure*}[htb]
	\centering
	\vspace{-0.25cm}
	\includegraphics[width=15cm]{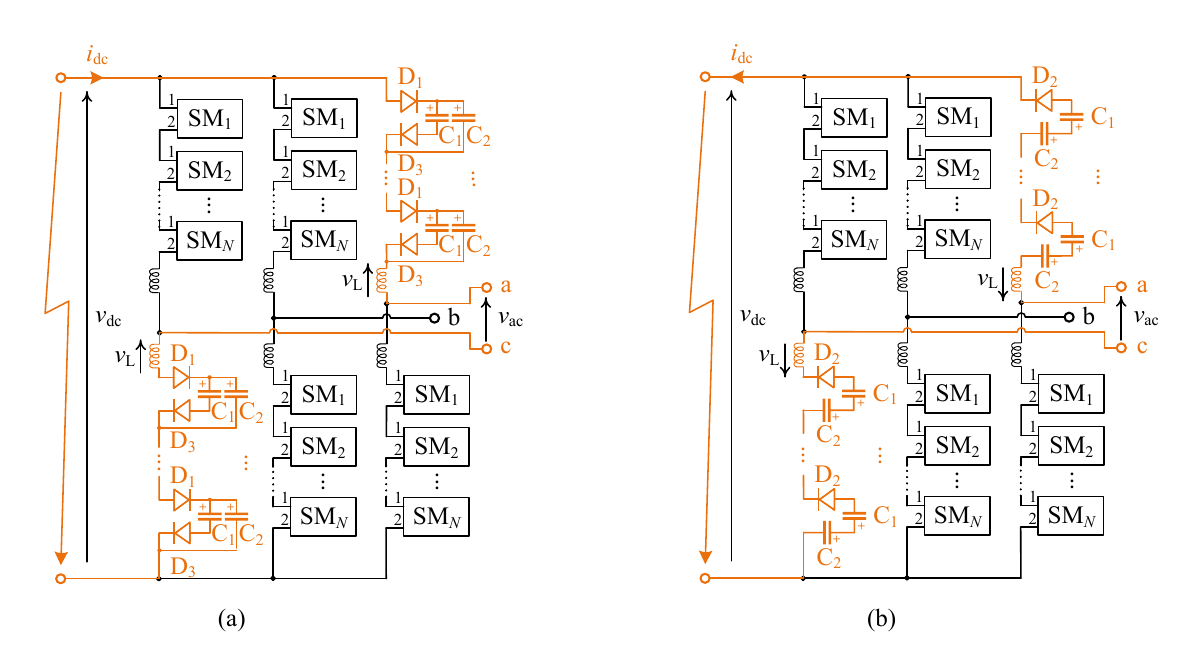}
	\vspace{-0.75cm}
	\caption{Current path of MMC during fault clearance : (a) $i_{\rm dc} > 0$; (b) $i_{\rm dc} < 0$.}
	\label{Fault_Current}
\end{figure*}

After the DC fault has been detected, all SMs are set to Mode 3. Depending on the direction of the arm-currents, two scenarios are analyzed. For simplicity without loss of generality, only one upper- and lower-arm are investigated here due to the system symmetry. When $i_{\rm dc} > 0$, a current path through phase A and C in the MMC is shown in Fig. \ref{Fault_Current}(a) as an example. The fault current flows through 2$N$ number of $\rm D_1$, $\rm D_3$, $\rm C_1$, and $\rm C_2$. According to the Kirchhoff's voltage law (KVL), following equation can be formulated:
\begin{equation}
	\label{KVL_1}
	2N v_{\rm D1}(t) + \sum_{}^{2N} v_{\rm c2}(t) + 2 v_{\rm L}(t) + v_{\rm ac}(t) - v_{\rm dc}(t) = 0, 
\end{equation}
or:
\begin{equation}
	\label{KVL_2}
	2N\left[ v_{\rm D1}(t) + v_{\rm D3}(t)\right] + \sum_{}^{2N} v_{\rm c1}(t)  + 2 v_{\rm L}(t) + v_{\rm ac}(t) - v_{\rm dc}(t) = 0,
\end{equation}
where $v_{\rm ac}(t)$ is the line voltage at the AC-side of the MMC, $v_{\rm D1}(t)$ and $v_{\rm D3}(t)$ are the voltages across diodes $\rm D_1$ and $\rm D_3$, respectively, and $v_{\rm dc}(t)$ is the residual DC-link voltage after the fault. 

Before the fault occurance, the MMC operates in steady state, and $\sum_{}^{N} v_{\rm c2}(t) = U_{\rm dc}$ if the voltage drops across the diode $\rm D_1$ are neglected, where $U_{\rm dc}$ is the rated DC-link voltage. Therefore, $\sum_{}^{2N} v_{\rm c2}(t) = 2U_{\rm dc}$, and \cref{KVL_1} can be rewritten as:
\begin{equation}
	\label{KVL_1r}
	v_{\rm D1}(t) = \frac{1}{2N} \left[ v_{\rm dc}(t) - 2 U_{\rm dc} - 2 v_{\rm L}(t) - v_{\rm ac}(t) \right]. 
\end{equation}
Similarly, when the voltage drops across the diodes $\rm D_1$ and $\rm D_3$ are neglected, $\sum_{}^{N} v_{\rm c1}(t) = U_{\rm dc}$ in the steady state. Therefore, \cref{KVL_2} can be rewritten as:
\begin{equation}
	\label{KVL_2r}
	v_{\rm D1}(t) + v_{\rm D3}(t) = \frac{1}{2N} \left[ v_{\rm dc}(t) - 2U_{\rm dc} - 2 v_{\rm L}(t) - v_{\rm ac}(t) \right]. 
\end{equation}

During normal operation, $|v_{\rm ac}(t)| \le \frac{\sqrt{3}}{2}U_{\rm dc}$ \cite{Yazdani2010}. After the fault occurs at the DC-side, $v_{\rm dc}(t) \ll U_{\rm dc}$ and $v_{\rm L}(t) > 0$, resulting in $v_{\rm D1}(t) < 0$ and $v_{\rm D1}(t) + v_{\rm D3}(t) < 0$ according to \cref{KVL_1r} and \cref{KVL_2r}, respectively. This indicates that all diodes $\rm D_1$ and $\rm D_3$ are immediately reversely biased when the DC fault occurs, and the fault current is blocked.

When $i_{\rm dc} < 0$, the current path of the MMC is shown in Fig. \ref{Fault_Current}(b). The fault current flows through 2$N$ number of $\rm D_2$, $\rm C_1$, and $\rm C_2$. The KVL equation can be formulated as:
\begin{equation}
	\label{KVL_3}
	\sum_{}^{2N} \left[ v_{\rm c2}(t) - v_{\rm c1}(t) \right] - 2N v_{\rm D2}(t) - 2 v_{\rm L}(t) + v_{\rm ac}(t) - v_{\rm dc}(t) = 0, 
\end{equation}
which can thus be rearranged as follows:
\begin{align}
	\label{KVL_3r}
	v_{\rm D2}(t) &= \nonumber \\
	\frac{1}{2N} & \left\{ \sum_{}^{2N}\left[ v_{\rm c2}(t) - v_{\rm c1}(t) \right] - 2 v_{\rm L}(t) + v_{\rm ac}(t) - v_{\rm dc}(t) \right\}.
\end{align}

When the DC fault just occurs at $t = t_{\rm f}$, $v_{\rm c2}(t_{\rm f}) - v_{\rm c1}(t_{\rm f}) \approx 0$, $v_{\rm dc}(t_{\rm f}) \ll U_{\rm dc}$, and $v_{\rm L}(t_{\rm f}) > 0$. Since $v_{\rm ac}(t_{\rm f})$ can either be positive or negative, $v_{\rm D2}(t)$ can either be above or below zero. If $v_{\rm ac}(t_{\rm f}) - 2 v_{\rm L}(t_{\rm f}) - v_{\rm dc}(t_{\rm f}) < 2Nv_{\rm F}$, where $v_{\rm F}$ is the forward voltage drop of a diode, then $v_{\rm D2}(t_{\rm f}) < v_{\rm F}$, indicating that the diodes $\rm D_2$ are reversely biased, and the fault current is immediately cleared. If $v_{\rm ac}(t_{\rm f}) - 2 v_{\rm L}(t_{\rm f}) - v_{\rm dc}(t_{\rm f}) > 2Nv_{\rm F}$, the diodes $\rm D_2$ are conducting. Since $\rm C_1 \gg C_2$, $v_{\rm c2}$ decreases much quickly than $v_{\rm c1}$ increases as time evolves. This reduces $v_{\rm ac}(t_{\rm f}) - 2 v_{\rm L}(t_{\rm f}) - v_{\rm dc}(t_{\rm f})$ quickly below $2Nv_{\rm F}$, making the diodes reversely biased, and finally the fault current will be blocked. In both scenarios discussed above, the MMC can block the DC-fault current regardless its direction using the proposed SM.

\section{Experimental Verification} \label{experiment}
\begin{figure}[htb]
	\centering
	\includegraphics[width = 6.5 cm]{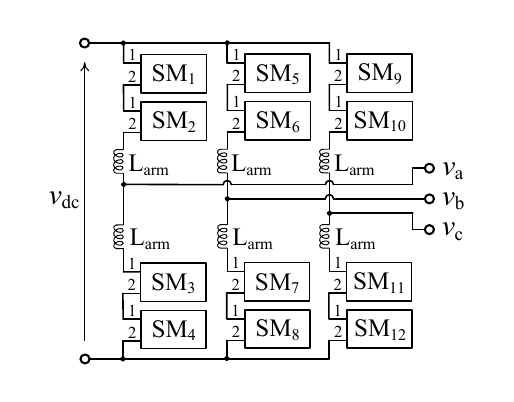}
	\caption{Circuit diagram of the MMC in the experiment.}
	\label{MMC_topology}
\end{figure}

\begin{figure}[htb]
	\centering
	\includegraphics[width = 8 cm]{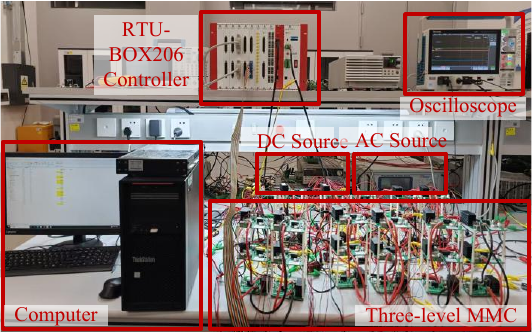}
	\caption{Experimental setup for the verification.}
	\label{Experiment_SetUp}
\end{figure}

The proposed SM topology is verified through an experiment in which the SM is applied to a three-phase MMC that can operate either as an inverter or a rectifier. The circuit topology of the MMC is depicted in Fig.~\ref{MMC_topology}, the experimental setup is shown in Fig.~\ref{Experiment_SetUp}, and the system parameters are listed in Table~\ref{Experiment_Parameters}. The main circuit is constructed using discrete components. The nearest level control algorithm (NLC) \cite{Tu2011} is adopted and implemented by an RTU-BOX206 controller from RTUnit to regulate the converter.
\begin{table}[ht]
	\caption{Parameters used in the experiment.}
	\label{Experiment_Parameters} 
	\centering
	\vspace{-0.25 cm}
	\begin{tabular}{p{2cm}<{\centering} @{} p{1cm}<{\centering} | @{} p{3.5cm}<{\centering} @{} p{1cm}<{\centering}}
		\hline \hline
		Parameter & Value & Parameter & Value \\
		\hline
		$\rm C_1$ (${\rm \mu F}$) & 470 & $\rm C_2$ (${\rm \mu F}$) & 47 \\
		$\rm L_{arm}$ (mH) & 1 & $\rm R_{load}$ ($\Omega$) & 10 \\
		$v_{\rm dc}$ (V) & 10 & $v_{\rm ac}$ (ph-ph, RMS, V) & 8.66 \\
		\hline \hline
	\end{tabular}
\end{table}

\begin{figure}[htb]
	\centering
	\includegraphics[width = 9 cm]{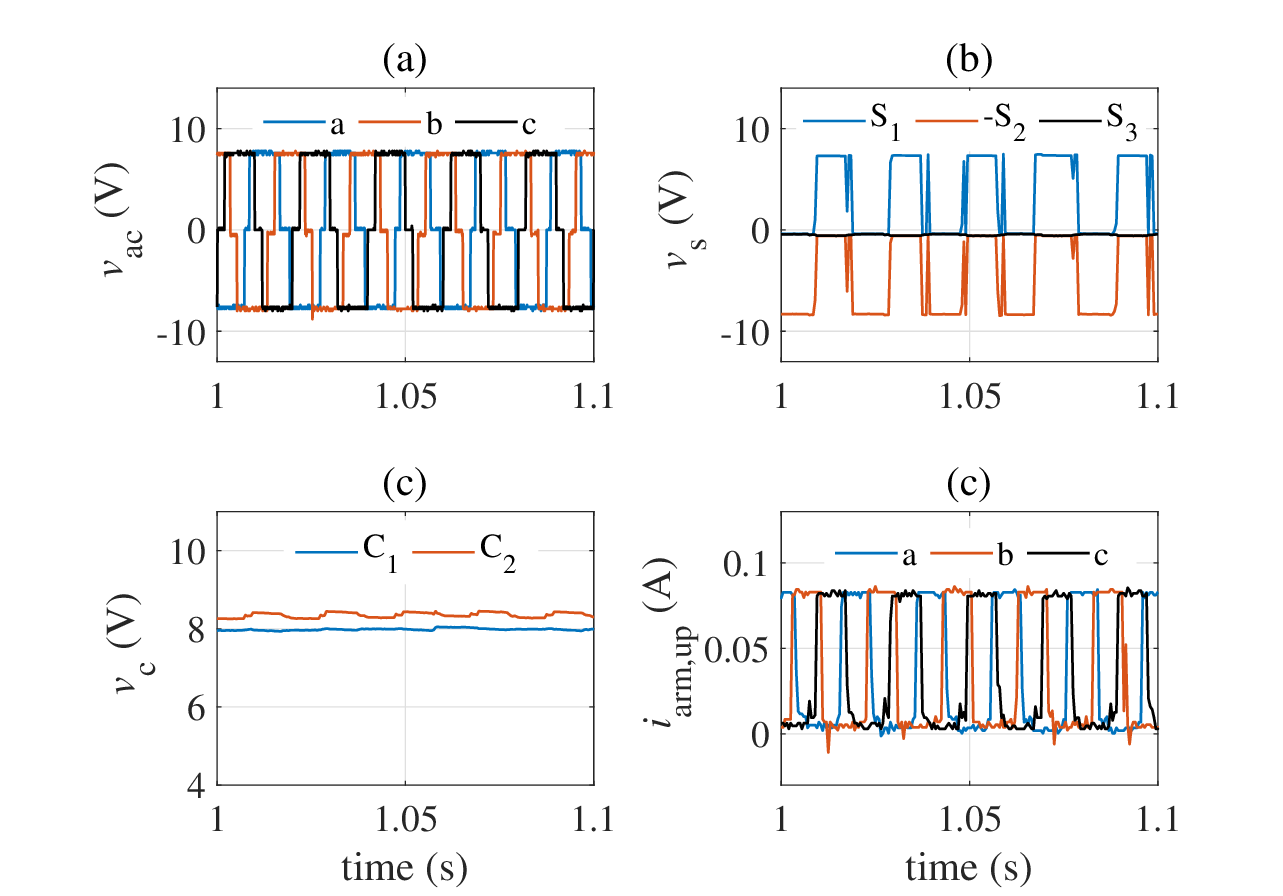}
	\caption{The MMC operating as inverter: (a) AC voltages; (b) voltage across the switches; (c) capacitor voltages; (d) upper arm currents.}
	\label{Inverter_experiment}
\end{figure}

When the MMC operates in inverter mode, its DC-side is powered by a DC voltage source with an output voltage of $v_{\mathrm{dc}}$, and each phase of the AC outputs is connected to a resistive load $\mathrm{R}_{\mathrm{load}}$. The waveform of the AC output voltages are plotted in Fig.~\ref{Inverter_experiment}(a). As can be observed from the figure, the MMC can generate 50 Hz three-phase three-level AC voltage using the proposed SM topology. During operation, the voltages of both capacitors $v_{\rm c1}$ and $v_{\rm c2}$ are around 8 V, and $v_{\rm c1}$ is slightly lower than $v_{\rm c2}$ due to the voltage drop over switch S$_3$, as shown in Fig.~\ref{Inverter_experiment}(c). The voltages across switches S$_1$ and S$_2$ are successively changed between the peak value at about 8 V and 0 V, indicating that they are successively turned-on and -off, as shown in Fig.~\ref{Inverter_experiment}(b). In contrast, the voltage across switch S$_3$ stays close to zero, suggesting that the switch is kept on during operation. 

\begin{figure}[htb]
	\centering
	\includegraphics[width = 9 cm]{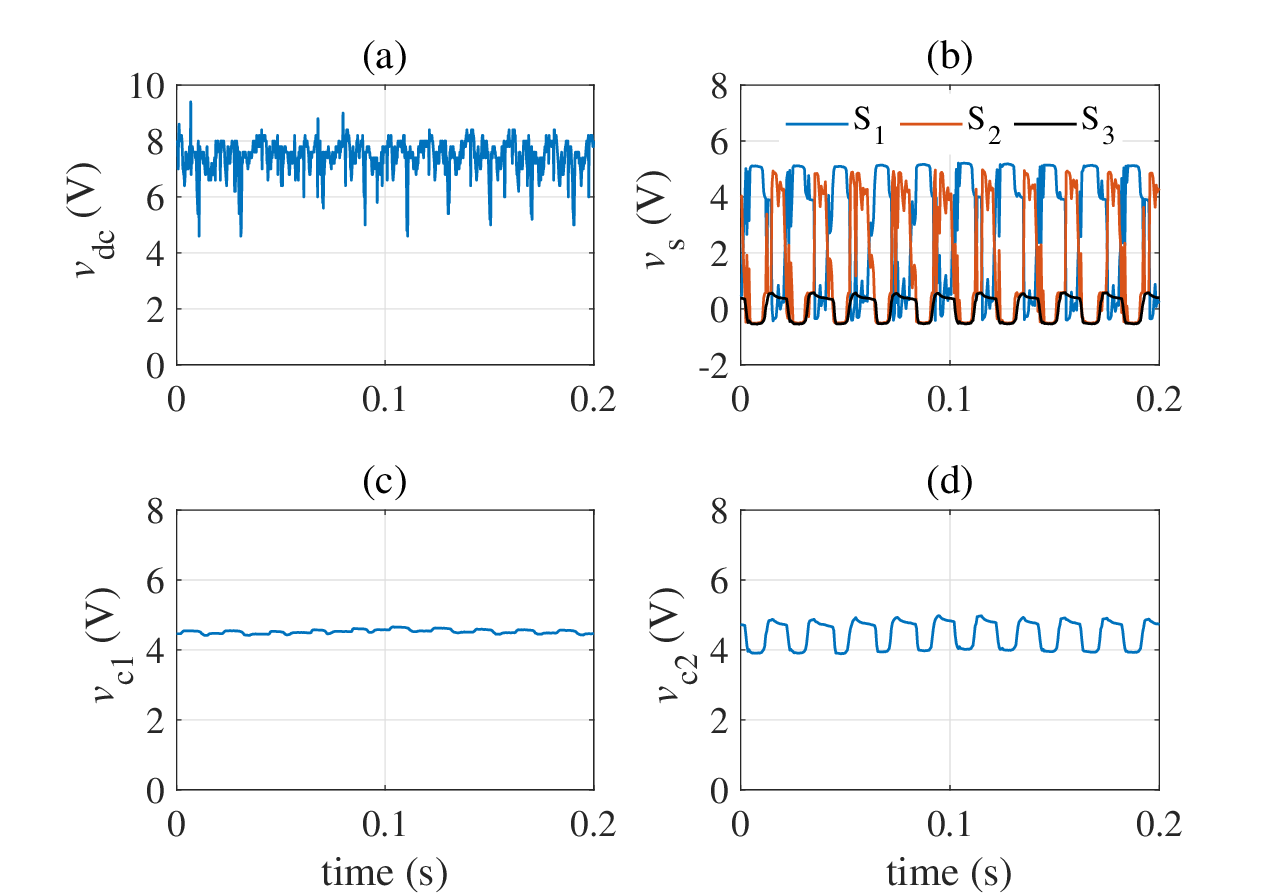}
	\caption{The MMC operating as inverter: (a) DC voltage; (b) voltage across the switches in SM$_{10}$; (c) voltage of C$_1$ in SM$_{10}$; (d) voltage of C$_2$ in SM$_{10}$.}
	\label{Rectifier_experiment}
\end{figure}

Conversely, when the MMC operates as rectifier, its AC ports are fed by a three-phase voltage source with phase voltages $v_{\mathrm{ac}}$. The DC output ports are connected to a resistive load $\mathrm{R}_{\mathrm{load}}$. Using the proposed SM, the MMC can convert AC power to DC power, as the waveform of DC output voltage shown in Fig.~\ref{Rectifier_experiment}(a). During rectification, the capacitor voltages $v_{\rm c1}$ and $v_{\rm c2}$ are regulated around $\frac{1}{2}v_{\rm dc}$, and $v_{\rm c2}$ oscillates greater than $v_{\rm c1}$ since C$_2$ is smaller than C$_1$, as shown in Fig.~\ref{Rectifier_experiment}(d) -- (d). Similar to the previous mode, switches S$_1$ and S$_2$ are successively turned-on and -off while switch S$_3$ is kept conducting, as shown in Fig.~\ref{Rectifier_experiment}(b).

\begin{figure}[htb]
	\centering
	\includegraphics[width = 9 cm]{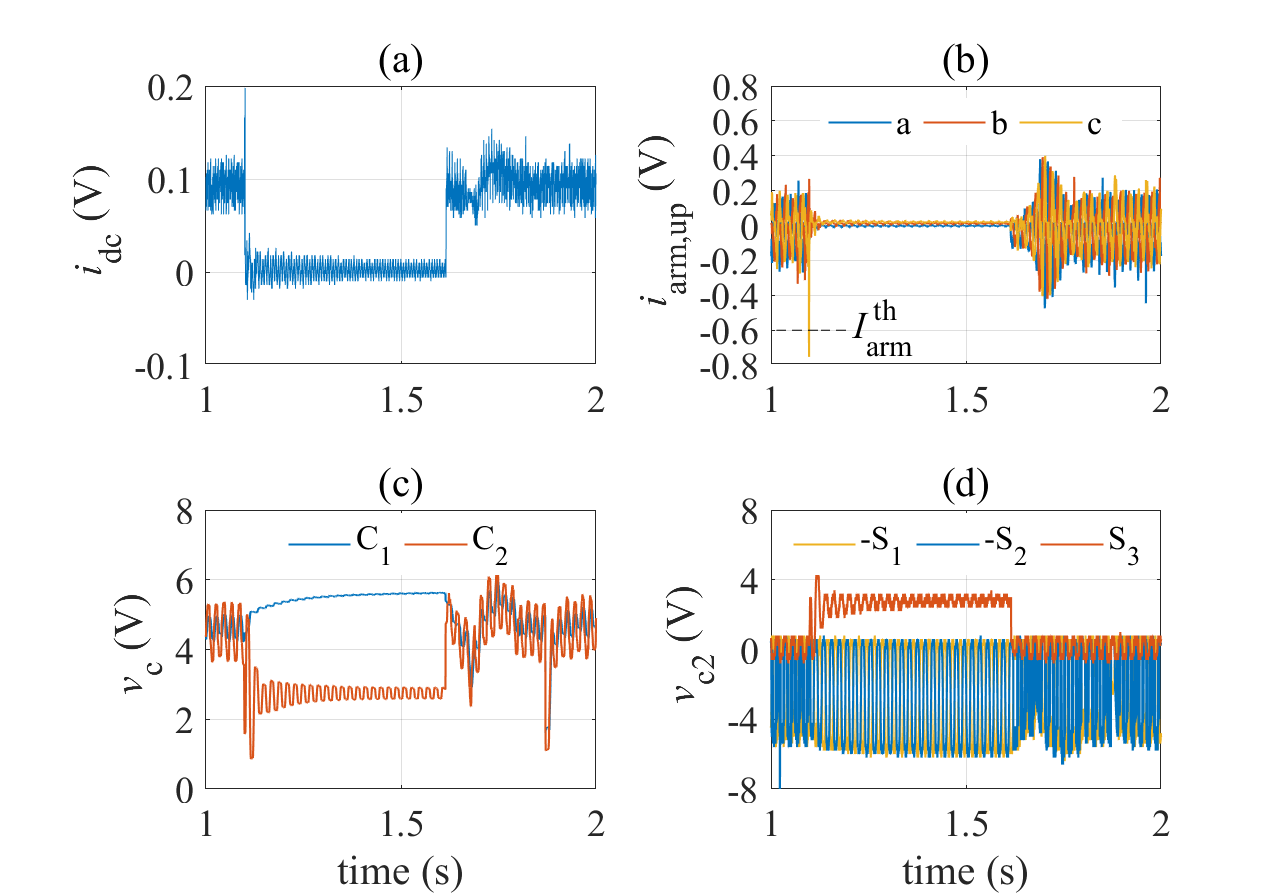}
	\caption{Experimental result of the MMC during non-permanent DC fault: (a) DC-current; (b) upper arm-currents; (c) capacitor voltage in SM$_{10}$; (d) switch voltages in SM$_{10}$.}
	\label{NonPermanent_DC_Fault}
\end{figure}

During rectification, a DC-side short-circuit fault occurs at about $t = 1.1$ s. Thereafter, the arm-current of the MMC increase rapidly as shown in Fig.~\ref{NonPermanent_DC_Fault}(b). Once the current exceeds a preset threshold value $I_{\mathrm{arm}}^{\mathrm{th}}$, which is set to $-0.3$ A in the experiment, the SMs are set to Mode 3 by turning off all switches. Consequently, both capacitors in each SM are connected in anti-series and inserted to the external circuit, as illustrated in Fig.~\ref{Operation_Blocked}(b). As discussed in the previous section, the arm-currents will be suppressed and blocked quickly in this case, as shown in Fig.~\ref{NonPermanent_DC_Fault}(b). Consequently, the DC-current falls quickly to zero, as shown in Fig.~\ref{NonPermanent_DC_Fault}(a). During the fault, $v_{\rm c1}$ increases below 6 V and $v_{\rm c2}$ decreases, as analyzed previously and depicted in Fig.~\ref{NonPermanent_DC_Fault}(c). As switch S$_3$ is turned-off during the fault, the voltage across it increases and is equal to the difference between $v_{\rm c1}$ and $v_{\rm c2}$, as can be seen in Fig.~\ref{NonPermanent_DC_Fault}(d). Meanwhile, the voltages across S$_1$ and S$_2$ become close to $v_{\rm ac}$. 


At approximately $t = 1.6$ s, the fault current is blocked and the fault has been extinguished, all SMs normal operation. The arm currents, AC-side currents, and capacitor voltages automatically return to their pre-fault levels, as shown in Fig.~\ref{NonPermanent_DC_Fault}. This experiment demonstrates that the proposed SM topology can block DC short-circuit fault without external protection devices and resume normal operation automatically after the fault is removed.

\section{Simulation} \label{simulation}
\begin{figure*}[h!]
	\centering
	\includegraphics[width = 13 cm]{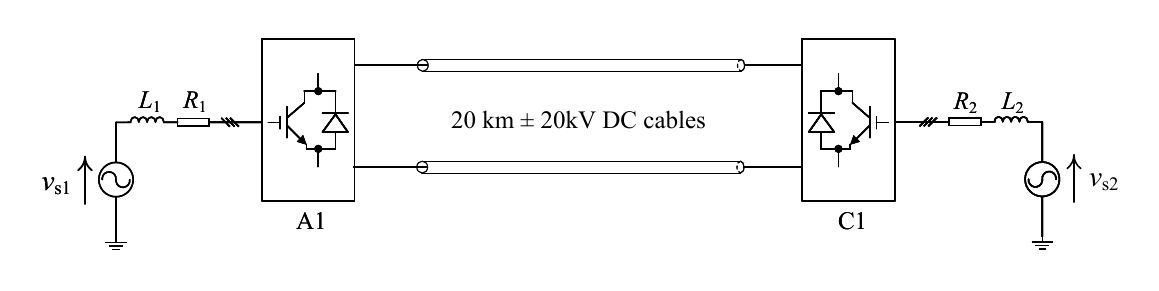}
	\caption{Topology of the HVDC-system used in the test simulation.}
	\label{Test_Simulation}
\end{figure*}

The proposed SM is tested in a high-voltage direct current (HVDC) transmission system rated at 80 MVA and $\pm 40$ kV with the topology as shown in Fig. \ref{Test_Simulation}, where the converter stations A1 and C1 are implemented as MMC using the proposed SM. The system is a $\frac{1}{10}$ scaled-down system of the DC system DCS1 in The CIGRE B4 DC Grid Test System \cite{Vrana2013}. The  transformer and the AC grid connected to the AC-side of converter A1 are modeled as an AC voltage source $v_{\rm s1}$, an inductor $L_1$, and a resistor $R_1$ connecting in series. Similarly, the AC voltage source $v_{\rm s2}$, the inductor $L_2$, and the resistor $R_2$ are used to model the transformer as well as the AC-grid connected to the AC-side of converter C1.

Each converter contains three upper- and lower-arms, and each upper- and lower-arm consists of 20 SMs as well as an arm inductor $L_{\rm arm}$ connecting in series. The NLC modulation algorithm is adopted to control the SMs and balance their voltages. The DC cables are modeled using the $\pi$-model. The test system is modeled and simulated using the library SimPowerSystems in MATLAB/Simulink 2021a. The values of all parameters used in the simulation are given in Table \ref{Simulation_Parameters}. 

\begin{table}[ht]
	\caption{Parameters used in the test simulation.}
	\label{Simulation_Parameters} 
	\centering
	\vspace{-0.25 cm}
	\begin{tabular}{p{3cm}<{\centering} @{} p{1cm}<{\centering} | @{} p{3cm}<{\centering} @{} p{1cm}<{\centering}}
		\hline \hline
		Parameter & Value & Parameter & Value \\
		\hline
		$\rm L_1$ (mH) & 3.977 & $\rm R_1$ ($\Omega$) & 0.0587 \\
		$\rm L_2$ (mH) & 6.522 & $\rm R_2$ ($\Omega$) & 0.0657 \\
		$\rm L_{arm}$ (mH) & 2.89 & $\rm R_{arm}$ ($\Omega$) & 0.5 \\
		$\rm C_1$ (mF) & 13 & $\rm C_2$ (mF) & 1.3 \\
		$\rm L_{arm,A1}$ (mH) & 2.889 & $\rm R_{arm,A1}$ ($\Omega$) & 0.005 \\
		$\rm L_{arm,C1}$ (mH) & 2.889 & $\rm R_{arm,C1}$ ($\Omega$) & 0.005 \\
		$v_{\rm s1}$ (ph-ph, RMS, kV) & 22 & $v_{\rm s2}$ (ph-ph, RMS, kV) & 22 \\
		Rating $v_{\rm dc}$ (kV) & $\pm$20 & $N_{\rm SM}$ & 20 \\
		Rating power (MVA) & 80 & Cable length (km) & 20 \\
		$\rm L_{cable}$ (mH/km) & 2.112 & $\rm R_{cable}$ ($\Omega$/km) & 0.0095 \\
		$\rm C_{cable}$ ($\rm \mu$F/km) & 0.2104 & $\rm G_{cable}$ ($\rm \mu$S/km) & 0.0620 \\
		\hline \hline
	\end{tabular}
\end{table}

\begin{figure}[h!]
	\centering
	\includegraphics[width = 9 cm]{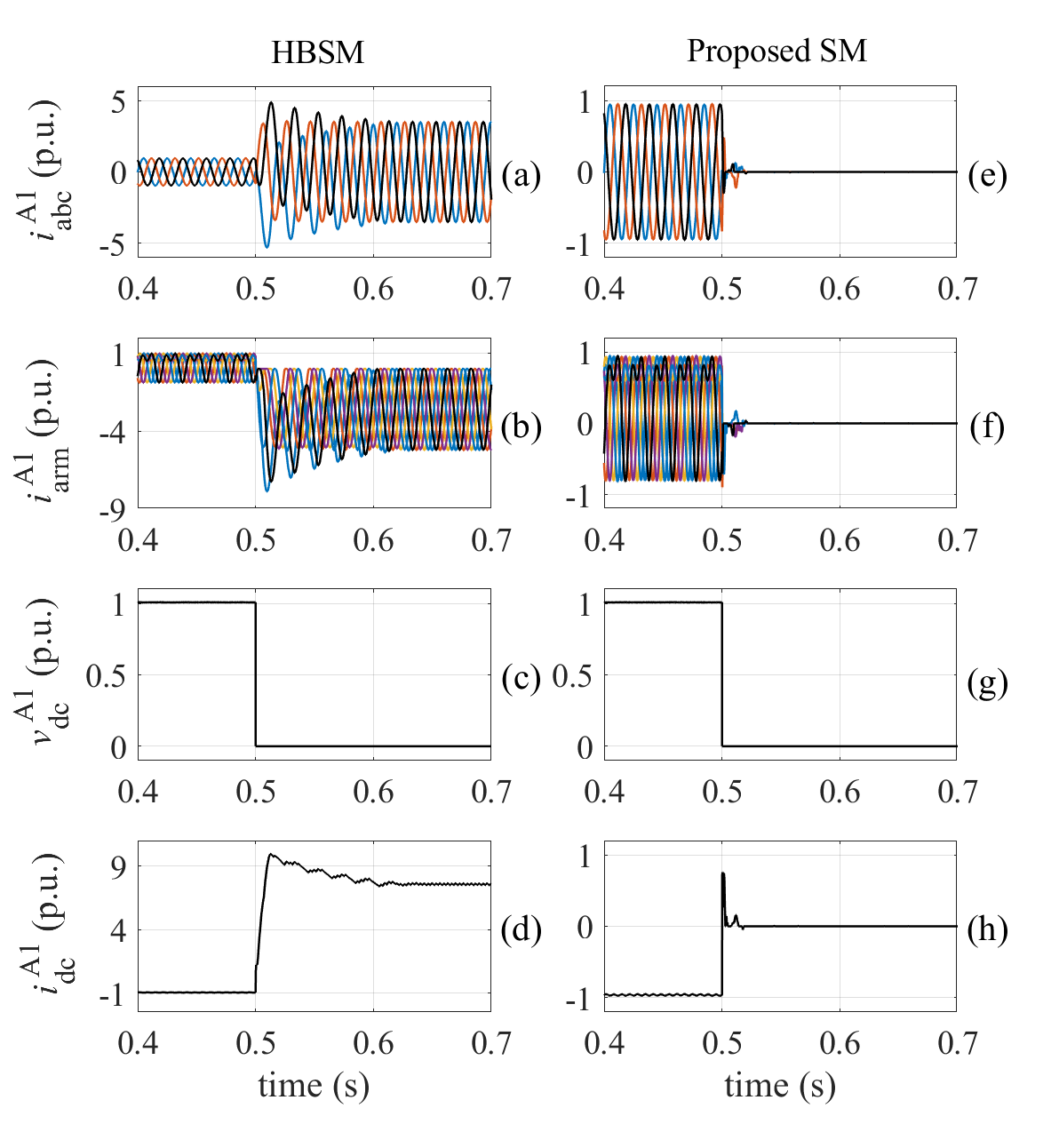}
	\caption{Comparison of response of MMC A1 using HBSM and the proposed SM during permanent DC fault. HBSM: (a) AC-grid currents; (b) arm currents; (c) DC-link voltage; (d) DC-link current. The proposed SM: (e) AC-grid currents; (f) arm currents; (g) DC-link voltage; (h) DC-link current.}
	\label{PermannentFault}
\end{figure}

Using the proposed SM, the converters A1 and C1 can operate stably in inverter mode and rectifier mode, respectively. Since both converters are set identical to each other, only the simulation results of A1 are plotted. The AC-side currents of both converters A1 and C1 are sinusoidal with constant amplitude in 1 per-unit (p.u.), as the waveform before $t = 0.5$ s shown in Fig.\ref{PermannentFault} (e). Similarly, the arm currents, the DC-side voltage, and the DC-side current have reached steady state at $t = 0.5$ s, as shown in Fig.\ref{PermannentFault} (f) -- (h). For comparison, the simulation results of the same test system using the HBSM are shown in Fig.\ref{PermannentFault} (a) -- (d) as reference. 

Two scenarios are considered and analyzed in the test simulation. In the first scenario, a permanent DC fault occurs at $t = 0.5$ s. In the second scenario, the DC fault is non-permanent and is cleared after a short period of time.  

\subsection{Scenario 1: Permanent DC Fault}
After the DC fault occurs, the DC-link voltage falls immediately to zero, and the DC-side current of A1 quickly changes its direction toward the external fault, as depicted in Fig. \ref{PermannentFault} (c), (d), (g), and (h). Once the DC fault is detected, all three switches in the proposed SM are switched-off. Thereafter, the output currents at the AC-side of A1 and the arm current within A1 quickly decay to zero using the proposed SM, as can be seen from Fig. \ref{PermannentFault} (e) and (f), respectively. In contrast, when the HBSM is adopted, these currents increase to a much higher level then normal operation, as shown in Fig. \ref{PermannentFault} (a) and (b), and these may cause system failure if no further protection measurements are taken. 

\begin{figure}[h!]
	\centering
	\includegraphics[width = 9 cm]{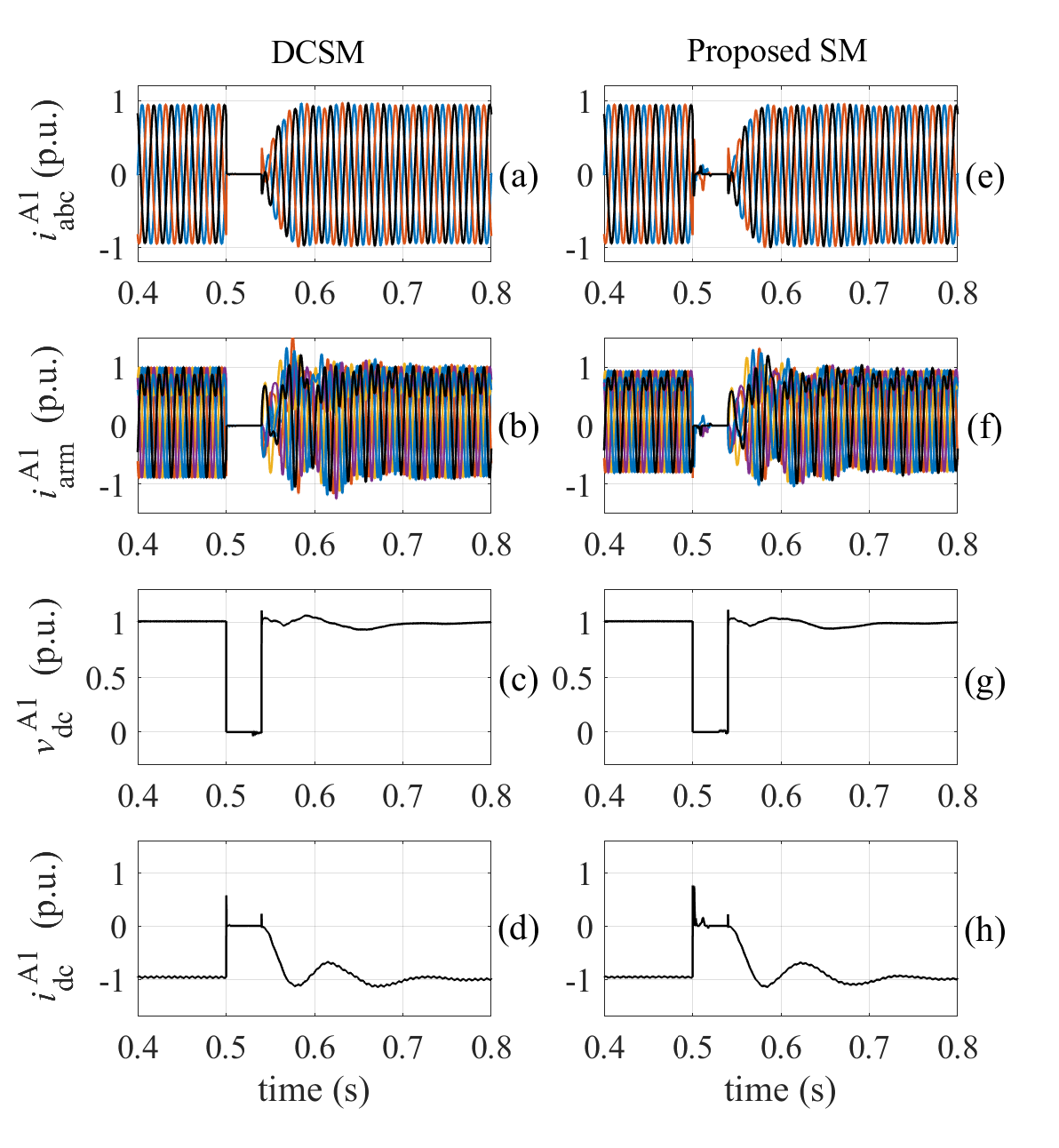}
	\caption{Comparison of response of MMC A1 using DCSM and the proposed SM during non-permanent DC fault. DCSM: (a) AC-grid currents; (b) arm currents; (c) DC-link voltage; (d) DC-link current. The proposed SM: (e) AC-grid currents; (f) arm currents; (g) DC-link voltage; (h) DC-link current.}
	\label{NonPermannentFault}
\end{figure}

\begin{figure}[h!]
	\centering
	\includegraphics[width = 9.5 cm]{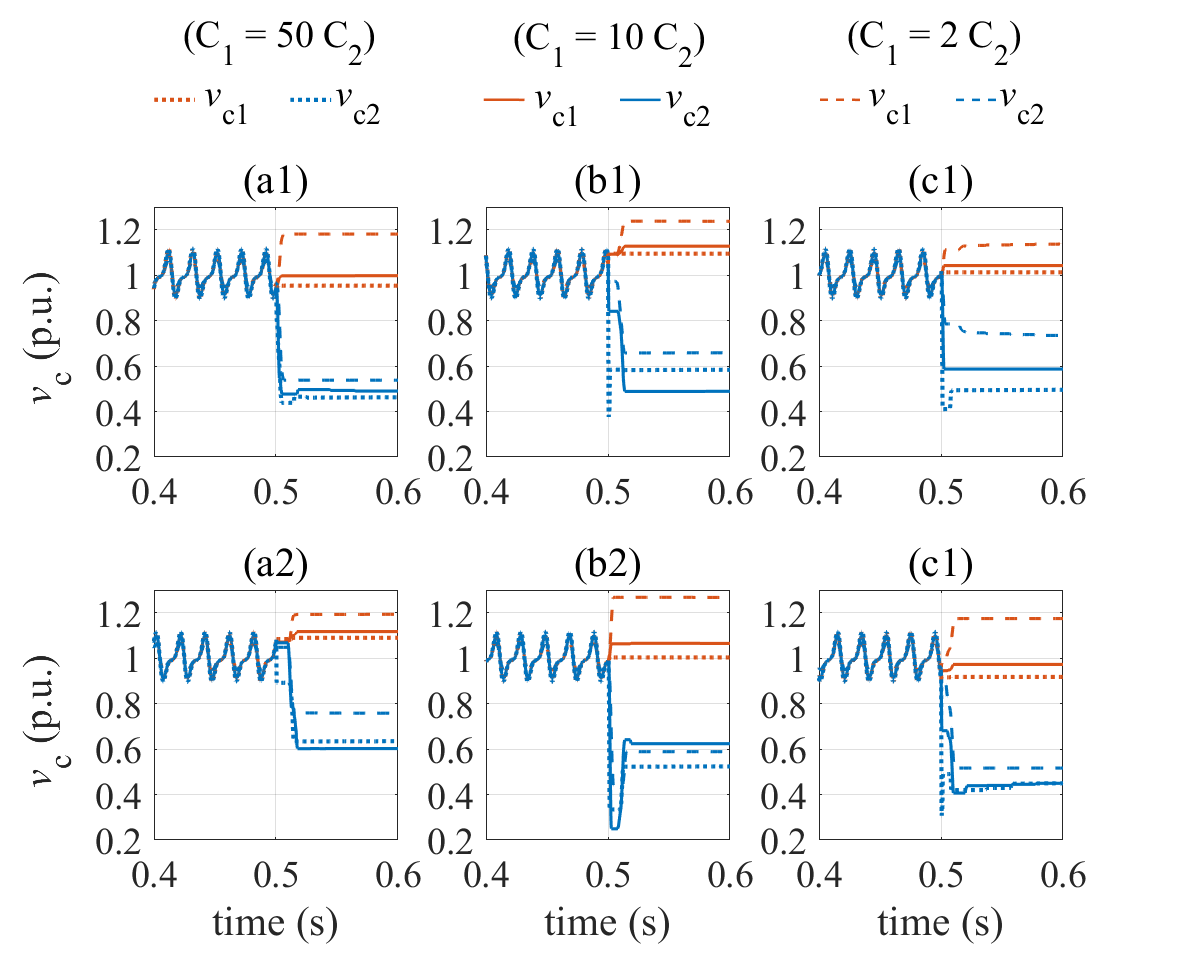}
	\caption{Voltage response of ALL capacitors in A1 during permanent DC fault in: (a1) upper arm of phase A; (a2) lower arm of phase A; (b1) upper arm of phase B; (b2) lower arm of phase B; (c1) upper arm of phase C; (c2) lower arm of phase C.}
	\label{Vc_PermannentFault}
\end{figure}

\begin{figure}[h!]
	\centering
	\includegraphics[width = 9.5 cm]{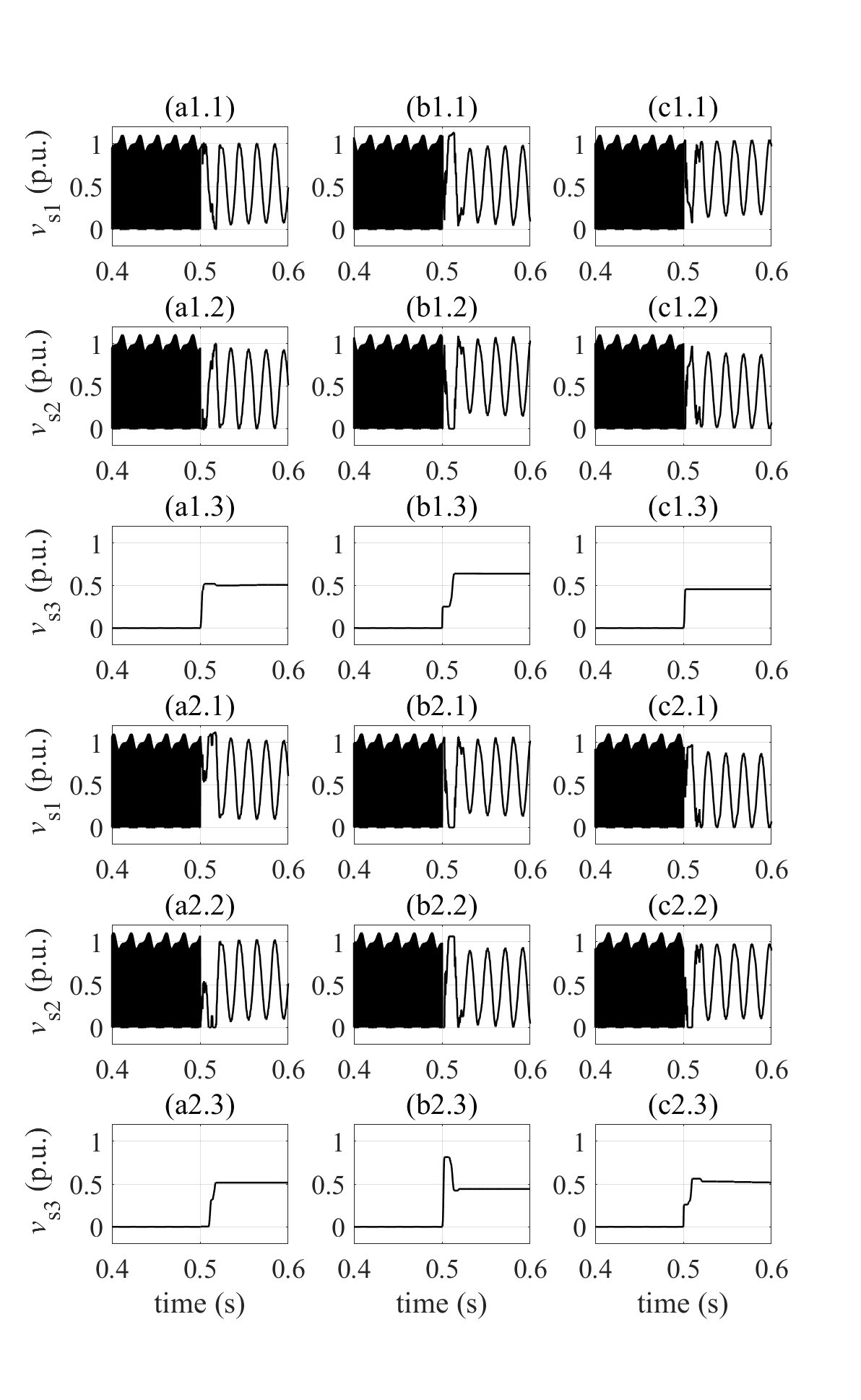}
	\caption{Voltages across the switches during permanent DC fault. Plot ($xi,j$) refers to all switches ${\rm S}_j$ in the upper arm ($i = 1$) or lower arm ($i = 2$) of phase $x$.}
	\label{Vs_PermannentFault}
\end{figure}

Meanwhile, the capacitor voltage of $\rm C_1$ in all SMs, $v_{\rm c1}$, increases slightly. Since $\rm C_1$ and $\rm C_2$ are now connected in anti-series and $\rm C_2 \ll C_2$, the capacitor voltage of $\rm C_2$ in all SMs, $v_{\rm c2}$, decreases more significantly compared to $v_{\rm c1}$, as shown in Fig. \ref{Vc_PermannentFault}. However, this transient ends up quickly, and both $v_{\rm c1}$ and $v_{\rm c2}$ become constant under the maximum value of 1.13 p.u.. The voltage across the switches $\rm S_1$ and $\rm S_2$ in all SMs, $v_{\rm s1}$ and $v_{\rm s2}$, respectively, are within the range of normal operation, as shown in Fig. \ref{Vs_PermannentFault}. The voltage $v_{\rm s3}$ across the switch $\rm S_3$ in all SMs increases after the fault occurrence because $\rm S_3$ are switched-off during the fault, as can be seen in Fig. \ref{Vs_PermannentFault}. However, the maximum value of $v_{\rm s3}$ is still under 1 p.u.. The simulation results suggest that the proposed SM can suppress the fault currents without causing significant internal over-voltage for the capacitors and the switches. 
\begin{figure}[h!]
	\centering
	\includegraphics[width = 9.5 cm]{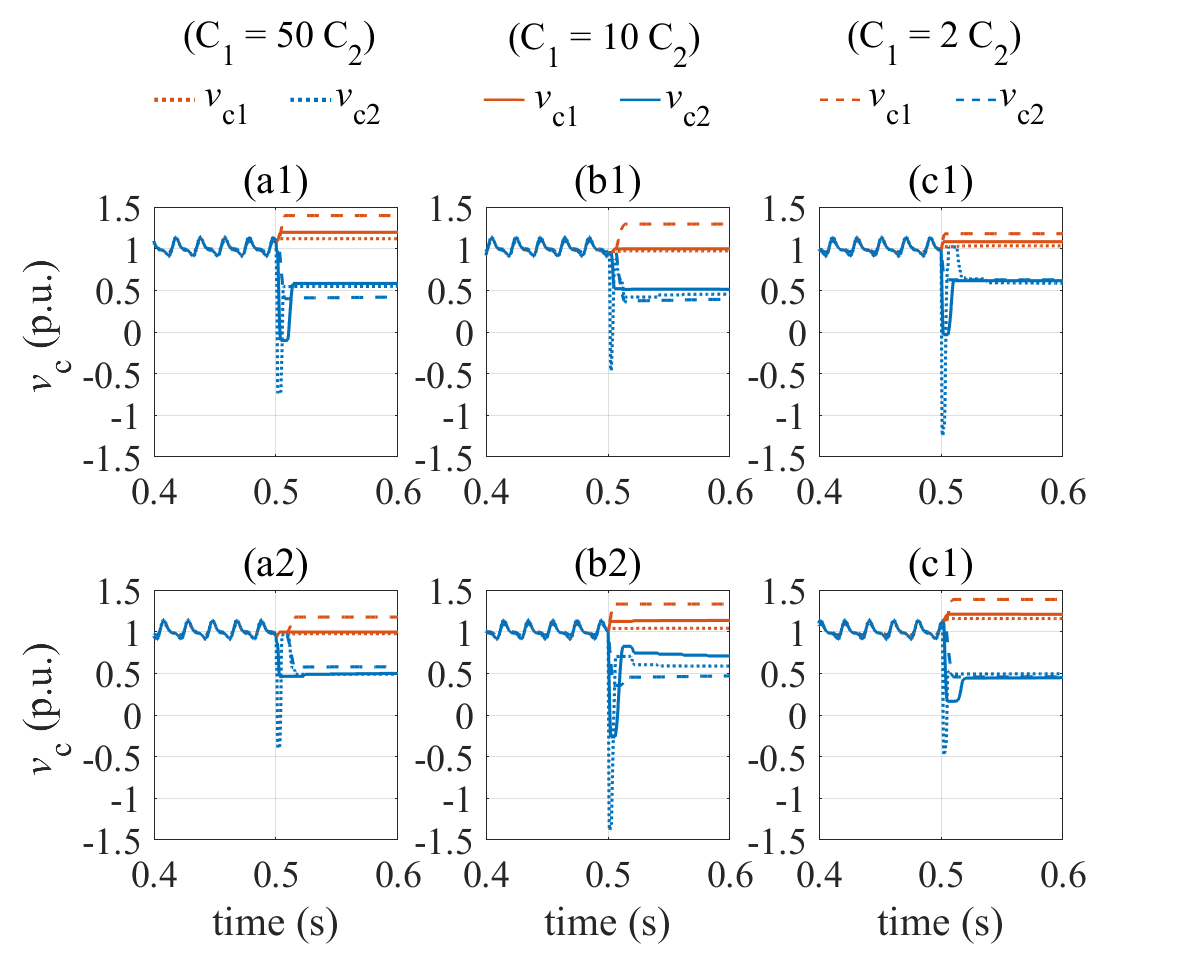}
	\caption{Voltage response of ALL capacitors in C1 during permanent DC fault in: (a1) upper arm of phase A; (a2) lower arm of phase A; (b1) upper arm of phase B; (b2) lower arm of phase B; (c1) upper arm of phase C; (c2) lower arm of phase C.}
	\label{Vc_PermannentFault_C1}
\end{figure}

\subsection{Scenario 2: Non-Permanent DC Fault}
\begin{figure}[h!]
	\centering
	\includegraphics[width = 9.5 cm]{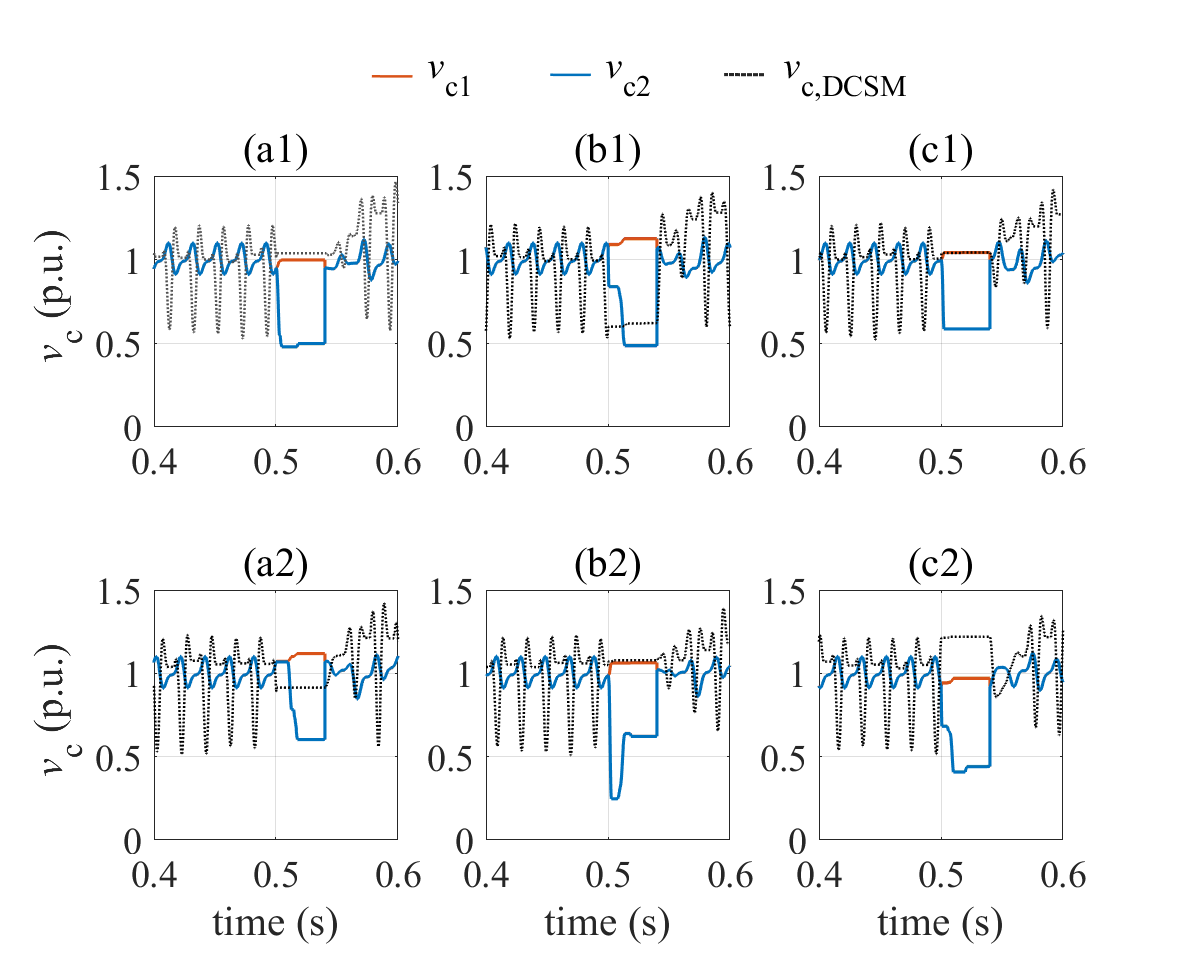}
	\caption{Voltage response of ALL capacitors during permanent DC fault in: (a1) upper arm of phase A; (a2) lower arm of phase A; (b1) upper arm of phase B; (b2) lower arm of phase B; (c1) upper arm of phase C; (c2) lower arm of phase C.}
	\label{Vc_NonPermannentFault}
\end{figure}

For the non-permanent DC fault occurred at $t = 0.5$ s, all switches are also turned-off as long as the fault is detected, the DC-link voltage and current, the AC currents, and the arm currents of A1 response similarly to the previous scenario as plotted in Fig.\ref{NonPermannentFault}. After the fault is cleared and the fault currents have been distinguished, all SMs are reset back to normal operation modes at $t = 0.54$ s. Thereafter, the DC-link voltage and current, the AC- and the arm-currents are quickly restored to the level of normal operation, as can be seen from Fig.\ref{NonPermannentFault} (e) -- (f), respectively. This capability is comparable to the DCSM, as can be seen from the simulation results shown in Fig. \ref{NonPermannentFault} (a) -- (d). However, in the DCSM used in the simulation, $\rm C_1 = C_2 = 26$ mF, which is twice the capacity of $\rm C_1 = 13$ mF used in the proposed SM. If the capacity of both $\rm C_1$ and $\rm C_2$ in the DCSM are reduced to the same as $\rm C_1$ in the proposed SM, the capacitor voltage $v_{\rm c}$ become more volatile, as the dotted lines shown in Fig. \ref{Vc_NonPermannentFault}.

\begin{figure}[h!]
	\centering
	\includegraphics[width = 9.5 cm]{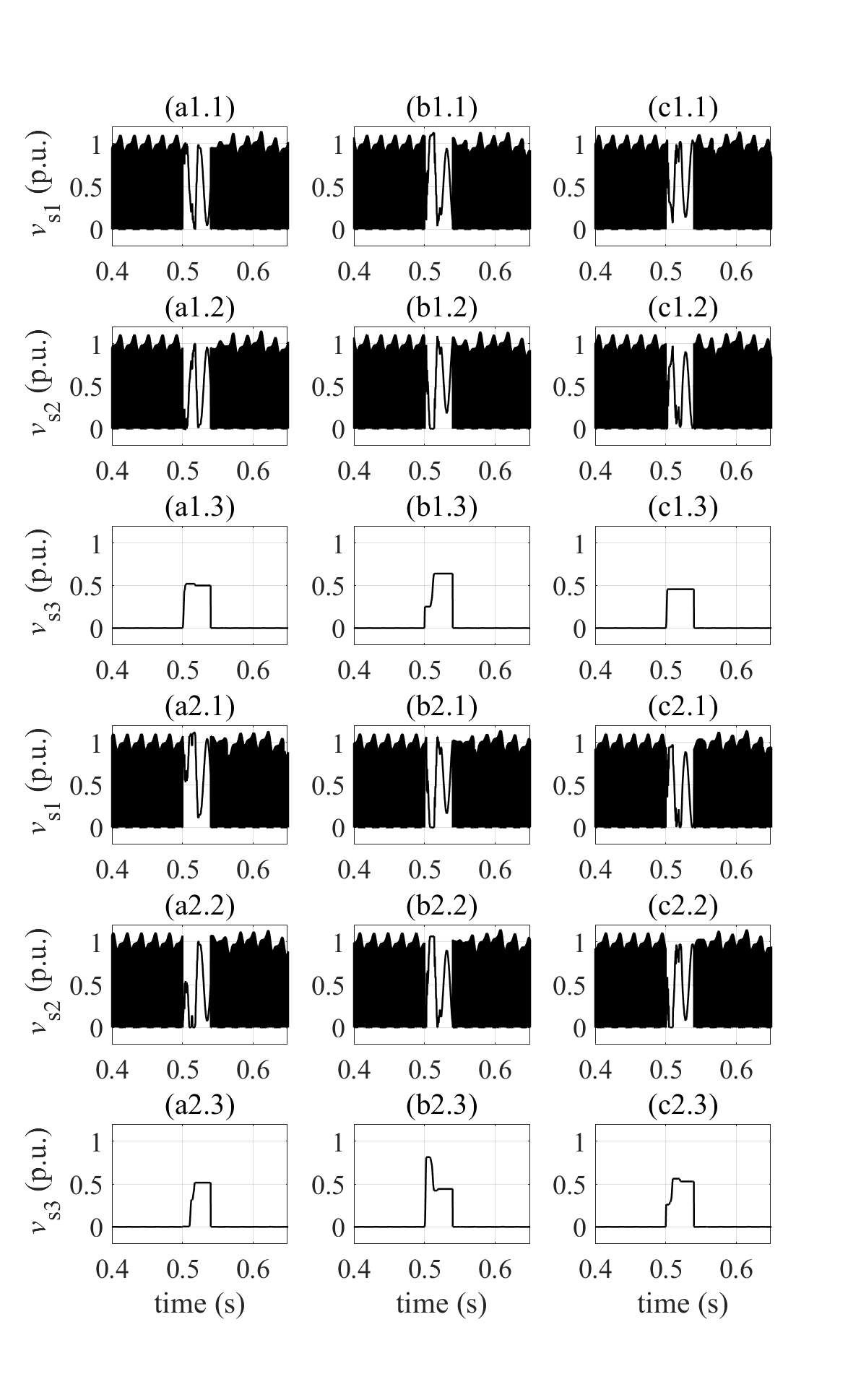}
	\caption{Voltages across the switches during non-permanent DC fault. Plot ($xi,j$) refers to all switches ${\rm S}_j$ in the upper arm ($i = 1$) or lower arm ($i = 2$) of phase $x$.}
	\label{Vs_NonPermannentFault}
\end{figure}

During the fault, the voltage across the capacitor $\rm C_1$ and $v_{\rm c1}$ in all SMs response similarly as in the previous case, as shown in Fig. \ref{Vc_NonPermannentFault}. After the fault has been cleared, the capacitor voltages in all SMs are automatically restored to the level of normal operation, as shown in the figure. For the switches, no over-voltage is detected during the fault and the fault-clearance process, as can be observed from Fig. \ref{Vs_NonPermannentFault}. The above simulation results suggest that the proposed SM exhibits comparable DC-fault clearance capability as the DCSM at a lower demand of capacitor capacity.

\section{Conclusion} \label{conclusion}
This paper has presented an enhanced HBSM with DC fault ride-through capability for MMCs. The topology and operating modes of the proposed SM are described in detail, and its performance is experimentally validated. Similar to the DCSM, the proposed SM can autonomously suppress fault currents and resume normal operation after fault clearance without relying on additional external protection devices. Unlike the DCSM, however, the proposed SM requires a smaller total capacitance while offering greater energy storage capability, and each proposed SM uses one fewer diode than the DCSM. These advantages help reduce the system complexity and overall capital cost of the MMC.



 
%

\bibliography{IEEEabrv,References}

\vspace{-1cm}


\end{document}